# Astronomical detection of a radioactive molecule $^{26}$AlF in a remnant of an ancient explosion

**Tomasz Kamiński[1]*, Romuald Tylenda[2], Karl M. Menten[3], Amanda Karakas[4], Jan Martin Winters[5], Alexander A. Breier[6], Ka Tat Wong[5], Thomas F. Giesen[6], Nimesh A. Patel[1]**

[1]Harvard-Smithsonian Center for Astrophysics, MS 78, 60 Garden Street, Cambridge, MA 02138, USA
[2]Department for Astrophysics, N. Copernicus Astronomical Center, Rabiańska 8, 87-100, Toruń, Poland
[3]Max-Planck Institut für Radioastronomie, Auf dem Hügel 69, 53121 Bonn, Germany
[4]Monash Centre for Astrophysics, School of Physics and Astronomy, Monash University, VIC 3800, Australia
[5]IRAM, 300 rue de la Piscine, Domaine Universitaire de Grenoble, 38406, St. Martin d'Héres, France
[6]Laborastrophysik, Institut für Physik, Universität Kassel, Heinrich-Plett-Straße 40, Kassel, Germany
*Correspondence to: tkaminsk@cfa.harvard.edu

**Decades ago, γ-ray observatories identified diffuse Galactic emission at 1.809 MeV (*1-3*) originating from $\beta^+$ decays of an isotope of aluminium, $^{26}$Al, that has a mean-life time of 1.04 million years (*4*). Objects responsible for the production of this radioactive isotope have never been directly identified, owing to insufficient angular resolutions and sensitivities of the γ-ray observatories. Here, we report observations of millimetre-wave rotational lines of the isotopologue of aluminium monofluoride that contains the radioactive isotope ($^{26}$AlF). The emission is observed toward CK Vul which is thought to be a remnant of a stellar merger (*5-7*). Our constraints on the production of $^{26}$Al combined with the estimates on the merger rate make it unlikely that objects similar to CK Vul are major producers of Galactic $^{26}$Al. However, the observation may be a stepping stone for unambiguous identification of other Galactic sources of $^{26}$Al. Moreover, a high content of $^{26}$Al in the remnant indicates that prior to the merger, the CK Vul system contained at least one solar-mass star that evolved to the red giant branch.**

Historic records (*8,9*) show that CK Vul or *Nova 1670* underwent an unusual outburst in 1670-1672. It was similar to outbursts of objects known as *red novae* which erupt in a stellar merger after which they cool off to low temperatures (*10, 11*). In this cool phase they produce large amounts of molecular gas and dust. CK Vul was recently discovered to be associated with a significant amount of dust and molecular gas, as well (*12*). What distinguishes CK Vul, even among red novae, is a high abundance of isotopes that are rare in matter of normal cosmic composition (*5,12*). In particular, our discovery of $^{26}$AlF in four rotational transitions (Methods) is the first firm detection of a radioactive molecule in any astronomical object, although numerous attempts to detect $^{26}$AlF have been made in the past (*13-15*). The unstable nucleus of $^{26}$Al is virtually absent in solar-composition objects and has a modest abundance of $10^{-5}$ with respect to $^{27}$Al in the Galactic interstellar medium whereas in CK Vul it is only ~7 times less abundant than the stable isotope $^{27}$Al (Methods).

The molecular remnant of CK Vul was discovered at millimeter wavelengths in rotational emission lines from a large variety of molecules (*5,12*). Imaging has shown that CO emission region has an extent of ~13" and a morphology of bipolar lobes and a torus-like feature which all are located at a center of a much larger (71") bipolar optical nebula of recombining plasma (*5,16*). The main AlF emission is observed in a small region of a full-width at half maximum size of 1.80(±0.05)×0.84(±0.06) arcsec, with the major axis at a position angle of 60°(±1°), and centered close to the radio-continuum source of CK Vul (*16*). At a distance of 0.7 kpc (*17*), the maximum extent corresponds to an *e*-folding radius of 430 AU. The AlF emission appears as a pair of two axisymmetric and collimated streams emanating from the center of the remnant and heading towards the east-northern and west-southern walls of lobes seen in CO and continuum dust emission (Fig. 1). Additionally, our most sensitive observations trace weak $^{27}$AlF emission, at a level of 3% of the peak, within the lobes out to a radius of 5.5 arcsec. The emission lines have an intrinsic full width of ~140 km/s, smaller than that of most other species observed in this source (*12*). The north-eastern part of the AlF region contributes most to the redshifted emission and the opposite side dominates the blueshifted emission, consistent with the overall kinematics of the molecular remnant.

Amongst all the molecular species that have been mapped thus far in CK Vul (examples of which are shown in Methods Fig. M1; see also Methods Table M3), the AlF emission has a unique distribution. That we observe the radioactive molecule of $^{26}$AlF only in a small region of the remnant is likely a chemical effect related to the formation and destruction of AlF. Observations of the AlF molecule in circumstellar media are rare but suggest that AlF forms close to stellar photospheres, i.e. at relatively high densities (*18 ,19*). Shocks and dust sputtering were also considered as a source of AlF (*20*). None of the scenarios can be excluded for CK Vul. The synthesis of AlF is likely limited by the elemental abundance of fluorine, not aluminum (*18*). The remnant can therefore contain other atomic and molecular forms of aluminum, some possibly depleted into dust. Thus, our AlF observations constrain only a lower limit on the content of $^{26,27}$Al in CK Vul. A search of other likely molecular carriers of aluminum, e.g. of AlCl, AlO, AlOH, and AlCN, has been performed (*12*) but none has been detected, suggesting a small reservoir of Al-bearing molecules other than AlF. On the other hand, a contribution from atomic Al to the recombining nebula may be significant. There are currently no observations that could be used to verify whether atomic aluminum is present. Based on the excitation analysis presented in Methods, we derive a total mass of the observed $^{26}$Al of $(3.4\pm1.8)\times10^{24}$ g, equivalent to about a quarter of the mass of Pluto.

The $^{26}$Al isotope is produced in the Mg-Al cycle in hydrogen burning via the $^{25}$Mg($p$, $\gamma$)$^{26}$Al reaction which requires temperatures above $30 \cdot 10^6$ K (*21*). It is thought to be efficiently produced in a variety of stars, including: classical novae with O-Mg-Ne white dwarfs; Wolf-Rayet stars; core-collapse supernovae; and asymptotic-giant-branch (AGB) stars that experienced *hot bottom burning* (*3, 22*). The progenitor of CK Vul was neither of these objects (*5, 17, 23*). However, more ordinary low-mass stars can produce $^{26}$Al as well. The $^{26}$Al synthesis takes place on the red giant branch (RGB) when hydrogen is burnt in a shell surrounding a helium core. Our model simulations (*30*; Methods) show that the most favorable conditions for producing $^{26}$Al occur when a star develops a condensed degenerate core, i.e. for initial stellar masses 0.8-2.5 $M_{\odot}$. The $^{26}$Al isotope is then deposited in a narrow outermost layer of the He core (Fig. 2). In a single RGB star, envelope convection never reaches the He core and therefore there is no way to dredge $^{26}$Al up to the stellar surface (and disperse it into the circumstellar and interstellar media). However, if the star is in a binary system and collides with a companion, matter from interiors of both stars can be mixed and ejected into the circumstellar medium. In particular, if the companion has a condensed core, the $^{26}$Al-rich outer layers of the He core of the RGB primary can be disrupted and exposed to eventually form a remnant such as that of CK Vul. Only a small portion of the available $^{26}$Al would have to be dispersed to explain the observed mass of $^{26}$Al and the aluminum isotopic ratio measured in CK Vul. Our calculations show that stars of 0.8-2.5 $M_{\odot}$ store a few times $10^{27}$ g of $^{26}$Al in the outermost layers of the He core, i.e. a factor of 1000 more than that found in CK Vul. Given this result and other observational constrains, a merger of two low-mass stars with at least one being on the RGB is the most likely scenario to explain CK Vul. Population-synthesis studies indeed indicate that low-mass binaries evolving off the main sequence to the RGB (and with orbital periods of 1–30 days) have a high chance to merge (*24,25*).

The $^{26}$Al decays are followed by emission of energetic positrons which may be an important local ionization source in CK Vul. Following Glassgold (*26*), our results suggest a $^{26}$Al-induced ionization rate of $2.0 \cdot 10^{-16}$ $s^{-1}$ per H nucleus for CK Vul. This is a lower limit considering that the derived $^{26}$Al mass is a lower limit and we adopted the solar elemental abundance for an object where aluminum is likely enhanced (*12*). The derived rate is the same as the typical ionization rate by cosmic rays in the Galactic disk (*27*). The regions of strong emission in the two ions, $N_2H^+$ and $HCO^+$, which were observed in CK Vul simultaneously with AlF, are more extended than that of $^{26}$AlF (Methods Fig. M1) suggesting that additional ionization mechanisms must be active in the remnant. It is possible that atomic forms of the radioactive nuclide of Al extend and ionize the remnant beyond the region traced in $^{26}$AlF emission or that other radioactive species are present in the remnant.

From the intensity of the 1.8 MeV line, it was estimated that all Galactic sources produce 1-3 $M_{\odot}$ of $^{26}$Al every 1 million yr (*1, 2, 28*). With our estimates on the $^{26}$Al mass in CK Vul, one would need ~1100 mergers like CK Vul going off every year to explain the entire Galactic content of $^{26}$Al. This figure is unrealistic as current rates of red novae suggest 1–2 such energetic transients per decade (*25*) and the rates are probably even lower for eruptions more characteristic of CK Vul (*7*). On the other hand, if the mass of $^{26}$Al in CK Vul is underestimated by a factor of 1100 – e.g. by not accounting for $^{26}$Al present in the atomic phase, other molecules, and solids – objects like CK Vul may be important contributors to the Galactic production of this radioactive nuclide. More observations and realistic models of the ionization and chemical structure of the remnant are necessary to investigate this issue further.

The 1.8 MeV emission arising from $^{26}$Al decays is hardly absorbed by interstellar or circumstellar matter (*26*) and easily escapes from the compact $^{26}$AlF region, even though it is heavily obscured by dust and gas (Fig. 1). Using our $^{26}$AlF observational constraints as a lower limit on the $^{26}$Al content in CK Vul, we calculate that the $^{26}$Al decay line has a flux of $\gtrsim 1.6 \cdot 10^{-10}$ $cm^{-2}$ $s^{-1}$, much below the sensitivity limit of the contemporary *SPectrometer on INTEGRAL* (SPI; $\sim 10^{-5}$ $cm^{-2}$ $s^{-1}$ in a $10^6$ s integration) (*29*). At such low estimated flux, it will be challenging to detect the 1.8 MeV line from CK Vul and probably from any other single stellar source even with future more sensitive γ-ray instruments. The case of CK Vul illustrates, however, that millimeter-wave spectroscopy, utilizing Atacama Large Millimeter/submillimeter Array (ALMA) and Northern Extended Millimeter Array (NOEMA), can now be used to study Galactic sources of radioactive nuclides, provided they produce molecules. Modern interferometer arrays not only can detect but also spatially identify discrete objects which are actively enhancing the Galaxy in $^{26}$Al. Because observations of molecules yield relatively easily the isotopologue (and thus isotopic) ratios, not available through γ-ray observations, millimeter-wave spectroscopy has also the potential to better identify the nucleosynthesis processes that lead to the Galactic $^{26}$Al production.

**Acknowledgments**. We are grateful to the directors K. Schuster, P. Cox, S. Dougherty, T. van Zeeuw, R. Blundell, and T. Beasley for granting us discretionary time at NOEMA, ALMA, APEX, SMA, and JVLA. T.K. thanks L. Matrá for an introduction to MCMC methods. R.T. acknowledges a support from grant 2017/27/B/ST9/01128 financed by the Polish National Science Centre. A.A.B. and T.F.G. acknowledge funding through the DFG priority program 1573 (Physics of the Interstellar Medium) under grant GI 319/3-1 and GI 319/3-2, and the University of Kassel through P/1052 Programmlinie "Zukunft". K.T.W. acknowledges support from the International Max Planck Research School (IMPRS) for Astronomy and Astrophysics at the Universities of Bonn and Cologne and also by the Bonn-Cologne Graduate School of Physics and Astronomy (BCGS).

This study made use of Atacama Pathfinder Experiment (APEX), which is a collaboration between the Max-Planck-Institut für Radioastronomie, the European Southern Observatory, and Onsala Space Observatory. Part of APEX data were collected under the programs 095.F-9543(A) and 296.D-5009(A). This paper makes use of the following ALMA data: ADS/JAO.ALMA#2015.A.00013.S and #2017.A.00030.S. ALMA is a partnership of ESO (representing its member states), NSF (USA) and NINS (Japan), together with NRC (Canada) and NSC and ASIAA (Taiwan) and KASI (Republic of Korea), in cooperation with the Republic of Chile. The Joint ALMA Observatory is operated by ESO, AUI/NRAO and NAOJ. The National Radio Astronomy Observatory is a facility of the National Science Foundation operated under cooperative agreement by Associated Universities, Inc. The IRAM 30m observations were carried out under projects 183-14, 161-15, and D07-14 and these with NOEMA under W15BN, E15AE, S16AV, and E16AC. IRAM is supported by INSU/CNRS (France), MPG (Germany) and IGN (Spain). The IRAM observations were supported by funding from the European Commission Seventh Framework Programme (FP/2007-2013) under grant agreement No 283393 (RadioNet3).

**Author Contributions.** T.K. wrote the text and analysed the observations. A.A.B. and T.F.G prepared the spectroscopic data used in this study. J.M.W. prepared, executed, and calibrated the NOEMA observations. KT.W. prepared and reduced the JVLA observations. T.K. prepared and reduced the ALMA and all single-dish observations. N.P. prepared and calibrated the SMA observations. R.T. and A.K. ran stellar-evolution models. All authors contributed to the interpretation of the data.

**Fig. 1. Maps of molecular emission of $^{26,27}$AlF**. Panels **A-F** show images and contours of emission in different transitions of $^{27}$AlF (**A-C**) and $^{26}$AlF (**D-F**), as indicated in each panel. Black contour is drawn at half the peak flux and the dashed cyan contour represents emission at the 3σ level. The contours overlap in panels D and E. In panels **G-H**, we compare the AlF 6-5 contours (from panel C) to the map of continuum emission (G) and that of CO 3-2 (H). All the colour images show flux in different linear scales. The cross indicates the position of the radio source.

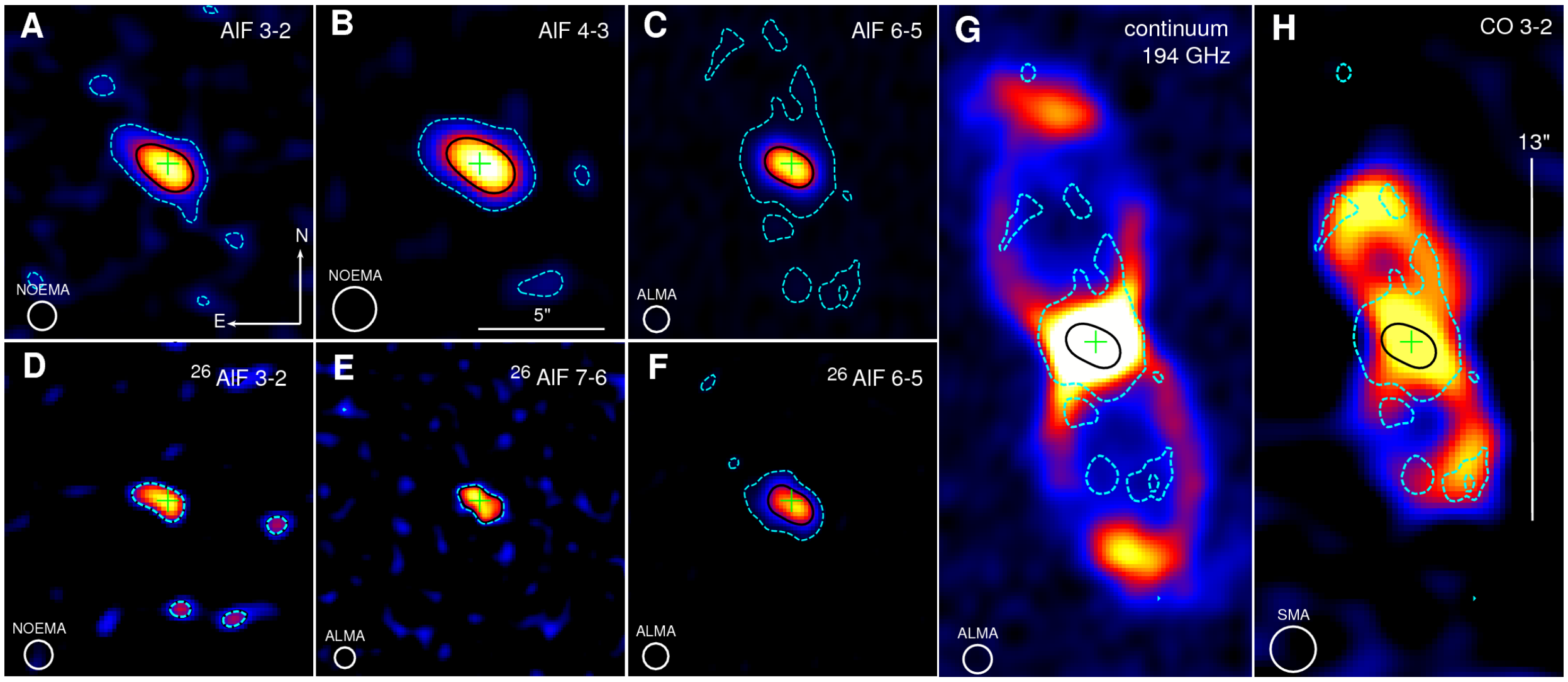

**Fig. 2.** Mass-abundance profiles of He, $^{26}$Al, and $^{27}$Al in a model of a 1 $M_\odot$ star at the tip of the RGB. The abundance of He (blue line) defines the extent of the He core and the H envelope which are indicated in the plot. The abundances of $^{26}$Al (red shaded area) $^{27}$Al (green dashed line) were scaled by a factor of $10^4$.

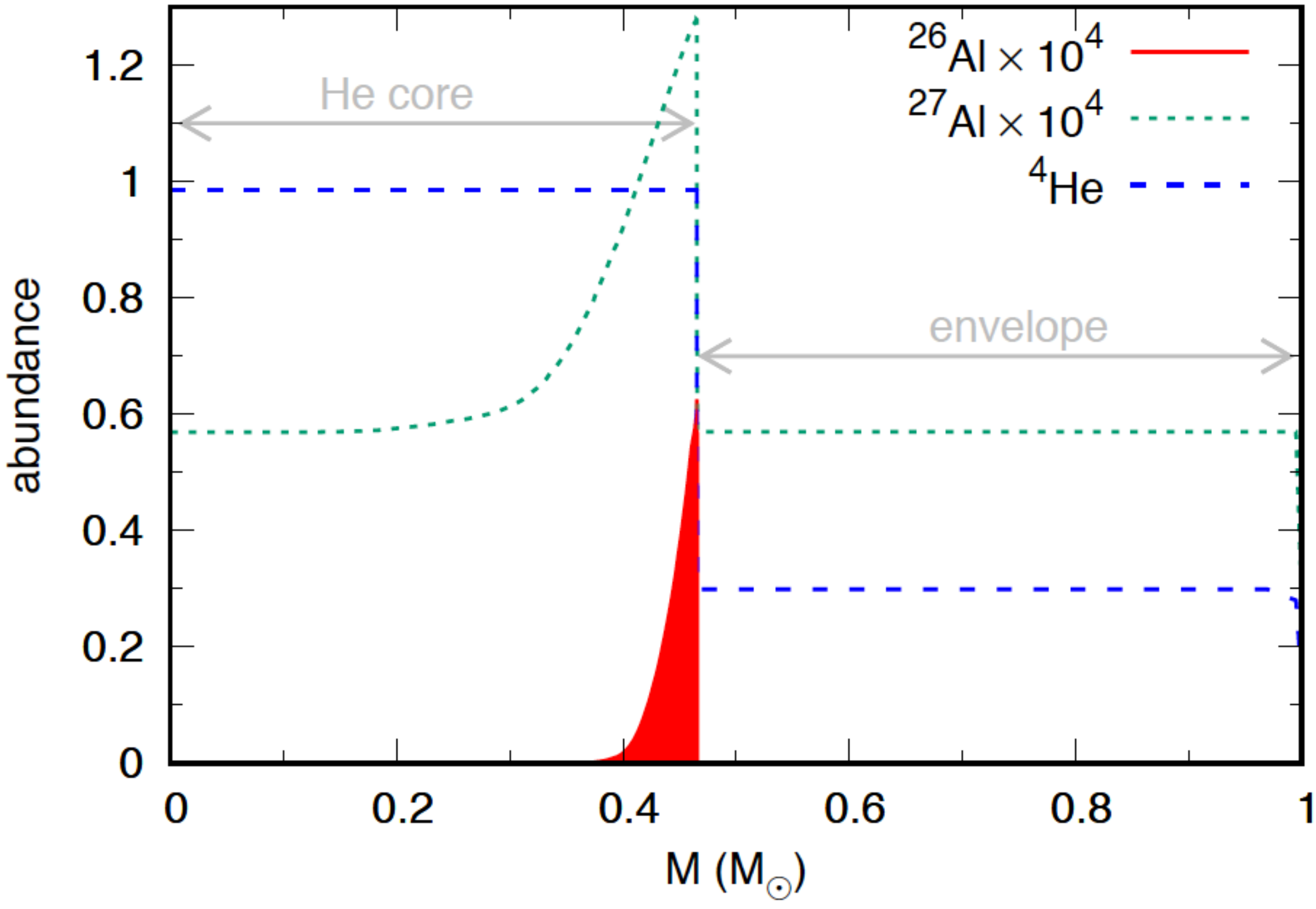

## METHODS

**Spectroscopic data for AlF isotopologues.** The identification and analysis of the pure rotational emission of $^{26}$AlF and $^{27}$AlF was based on spectroscopic data prepared in this study. For $^{27}$AlF, we used mass-scaled Dunham parameters and hyperfine constants derived from laboratory measurements (*31*). Accurate line positions of $^{26}$AlF were calculated through the mass-scaled Dunham parameters of $^{27}$AlF. We used fourth-order correction terms to derive positions of hyperfine components with an accuracy better than 1 MHz. The hyperfine splitting of $^{26}$AlF is more complex than that of $^{27}$AlF owing to a twice larger nuclear spin ($I$=5). To derive the hyperfine structure for $^{26}$AlF, we used higher-order Dunham corrections and scaled accordingly the $^{26}$Al electric quadrupole moment $Q$ (*ref. 32*) and the magnetic coupling parameter $c_I$ with the nuclear $g_N$ factor. A permanent dipole moment of $\mu$=1.53 D (*ref. 33*) was adopted for both isotopologues. The spectroscopic constants we used to generate the line lists are given in Table M1. Line frequencies, energies of the rotational levels above the ground ($E_u$), line strengths ($S_{ul}\mu^2$), and partition functions were derived using *pgopher* (*34*). The method of our calculations is similar to that used in earlier studies of rotational spectra of $^{26}$AlF (*14*).

Spectroscopic laboratory studies of rare radioactive materials such as $^{26}$AlF would be very challenging. Although laboratory measurements are usually needed to unambiguously identify complex molecules, it is not necessary for simple di-atomic species, especially for most astronomical applications. For di-atomic molecules and within the Born-Oppenheimer approximation, the mass dependence of spectra can be separated from that of the bond length. The underlying theory was developed by Dunham in 1932 (*43*) and has been successfully applied to many molecules. The spectra of diatomic molecules of identical bond length differ only in the mass scaling factors which are known to a high accuracy. Measurements for one isotopic species, e.g. $^{27}$Al$^{19}$F, are thus sufficient to determine spectra of other isotopologues, e.g. $^{26}$Al$^{19}$F. High-precision measurements of different isotopologues can show how accurate this approximation is and higher order corrections, if necessary, can be added. The corrections are often insignificant, in particular they are typically much smaller than the accuracy of astronomical observations (*45*). Our calculations included these corrections but they turned out to be negligible in case of the studied source whose line widths are of ~140 km/s.

Because the AlF lines are so intrinsically broad in CK Vul, their hyperfine structure is unresolved in most of our observations. As shown in Fig. M1, the hyperfine splitting is highest in the $J$=1-0 transition which was however observed at insufficient sensitivity to reveal the hyperfine structure (and is degraded in resolution in the figure). For both isotopologues, the hyperfine splitting caused by the $^{19}$F nucleus is negligible. The full spectroscopic data which include the hyperfine splitting of $^{26}$Al$^{19}$F and uncertainties in line positions are available online (goo.gl/uMV1h9). In our analysis, we often used the intensity-weighted mean frequency and the sum of line strengths $S_{ul}\mu^2$ to represent the position and strength of a given transition, respectively. The centroid line frequencies and line strengths used in the analysis are given in Table M2.

**Single-dish observations.** The presence of $^{27}$AlF in CK Vul was revealed in the line survey obtained with the IRAM 30 m and Atacama Pathfinder Experiment 12 m (APEX) telescopes in 2014-2017 (*12*). The survey provided measurements of $^{27}$AlF emission in the following transitions: $J$=3-2, 4-3, 5-4, and 6-5. Additionally, the 7-6 transition was covered by IRAM 30m spectra but it overlaps with a much stronger line of CO $J$=2-1and its flux could not be reliably measured. Two other lines of $^{27}$AlF, i.e. $J$=8-7 and 9-8, were covered but not detected. Their upper 5$\sigma$ limits and fluxes of the other detected lines are given in Table M2.

Most of the corresponding transitions of $^{26}$AlF were also covered in the survey. We find a weak spectral feature at the expected location of $^{26}$AlF 3-2 but the emission line partially overlaps with a broader instrumental feature (*12*). The 4-3 line is also visible and it partially overlaps with the $J,K_a,K_c$=8,2,6-8,1,7 line of $SO_2$ (see below). Only rough flux constraints on the $^{26}$AlF 4-3 emission could be obtained from this blending feature. The $J$=5-4 transition of $^{26}$AlF overlaps with a much stronger line of H$^{13}$C$^{15}$N 2-1 and the flux contribution of $^{26}$AlF could not be assessed. Another covered line, $J$=7-6, appears to be clear of any contamination but is seen at a modest ~3$\sigma$ level. All other lines are below our sensitivity limits. The three $^{26}$AlF features, i.e. $J$=3-2, 4-3, and 7-6, are observed at a modest signal-to-noise ratio (S/N) and their observations alone constitute only a tentative indication of the presence of $^{26}$AlF in CK Vul. The three features, however, provided a strong incentive to repeat the observations with more sensitive interferometers.

**Interferometric observations.** The sensitive observations of both AlF isotopologues were obtained with JVLA, NOEMA, and ALMA. The interferometric observations are summarized in Table M3. The line fluxes are given in Table M2. The following transitions were observed:

- The ***J*=1-0** transitions of $^{26}$AlF and $^{27}$AlF were covered by a *Ka*-band spectrum obtained with the Karl G. Jansky Very Large Array (JVLA) in the DnC configuration. The lowest transition of AlF has a significant hyperfine splitting which is comparable to the intrinsic line width of AlF emission (Fig. 1A). This extra broadening makes the line peak intensity lower and harder to detect than for lines at higher frequencies. Only after smoothing the spectrum to a resolution comparable to the hyperfine splitting is the emission of $^{27}$AlF 1-0 apparent (Fig. M1A). This transition should be considered as only tentatively detected. The flux of the line is at a 2–5$\sigma$ level, which is insufficient to provide a good-quality map of the $^{27}$AlF 1-0 emission. In order to extract the spectrum, we used an aperture defined in maps of $^{27}$AlF 3-2 from NOEMA. The $^{26}$AlF 1-0 emission is not detected in the JVLA data, consistent with the isotopologue ratio derived in this study.
- The ***J*=3-2** transition of both AlF isotopologues was observed with the emerging NOEMA interferometer. Observations were obtained in 2016 and 2017 with 7 and 8 antennas, respectively. The WideX correlator was used. Both transitions are detected and their emission regions resolved with a beam of ~0.78 arcsec. At this angular resolution, the peak S/N of the two emission regions is 24 and 5 (and higher for source-averaged fluxes).
- The ***J*=4-3** transitions were observed with 6 antennas of NOEMA and with WideX. The $^{27}$AlF emission region is marginally resolved by the beam of ~1.64 arcsec. The $^{27}$AlF emission may be contaminated by the HNCO (6,0,6-5,0,5) line whose rest frequency is 29.7 km/s away from that of $^{27}$AlF 4-3. The separation is smaller than the full width at half maximum (FWHM) of the observed feature of 40.8 km/s. An excitation model of HNCO based on the single-dish survey (*12*) implies that less than 6% of the total flux of the observed feature may come from HNCO. However, the accuracy of this model is modest and the model does not take into account the potential difference in spatial distributions of HNCO and AlF emission. The characteristics of the emission region ascribed here to $^{27}$AlF 4-3 are the same as these of other AlF transitions and do not indicate any sign of contamination. We therefore neglect the potential contribution from HNCO and interpret the total flux of the emission feature as that of $^{27}$AlF 4-3. The corresponding $^{26}$AlF 4-3 transition is detected by NOEMA but blends partially with the $SO_2$ 8,2,6-8,1,7 line whose rest frequency is blueshifted by 55.9 km/s with respect to the $^{26}$AlF line. Interferometric ALMA imaging of another line of $SO_2$, 4,2,2-3,1,3, shows that the extent of the emission region of $SO_2$ is similar to that of AlF (Fig. 1) and therefore the relative contribution of the two species to the blend cannot be disentangled based on spatial information only. The kinematic separation of the $SO_2$ and $^{26}$AlF lines is however wide enough to perform a deblending procedure in which the best-fitting combination of two Gaussian components gives the line characteristics. The line centers were fixed while the amplitudes and widths were free parameters in this $\chi^2$ minimization procedure. The results are shown in Fig. M1 H-J and the flux of $^{26}$AlF 4-3 is given in Table M2. The flux of the $SO_2$ line is 2.34 times higher than that of $^{26}$AlF.
- Both isotopologues were observed with ALMA in the ***J*=6-5** lines located in ALMA Band 5. These are the most recent observations (April 2018) and provided us with spectra and maps of the best sensitivity. At a beam size of 1.4×1.1 arcsec (at natural weighting), both emission regions are resolved and their peaks are observed at S/Ns of 170 and 33 (the source-averaged fluxes give even higher S/Ns).
- The ***J*=7-6** transition of both isotopic species was observed with ALMA in Band 6. Both lines are observed at a high S/N and are well resolved with a 0.8 arcsec beam. The $^{27}$AlF line overlaps with a broad wing of an intense line of CO 2-1. The location of the $^{27}$AlF 7-6 line corresponds to a velocity of 333 km/s with respect to the center of the CO 2-1 emission. At this velocity, the CO emission is relatively faint and extended, i.e. most of the CO emission is spread over a region of a radius of ~6 arcsec from the center of the molecular remnant. Our maps of AlF transitions, including the map of $^{26}$AlF 7-6 from the same ALMA dataset, show that the AlF emission is enclosed within a radius of 1.05 arcsec. Hence, by extracting the signal within the AlF emission region defined in other observations, we minimized the contamination from the CO emission. Furthermore, the spectrum extracted within the AlF emission region reveals that the $^{27}$AlF 7-6 transition partially overlaps with an emission feature which we tentatively identify as a transition of methylamine, $CH_3NH_2$ 3,2,4-3,1,5, whose rest frequency is 140 km/s away from that of $^{27}$AlF 7-6. To measure the pure flux of $^{27}$AlF 7-6, we replaced the small contaminated part of its original profile by the mirrored unaffected wing, as shown in Fig. M1 (panel F). A flux measured in such a modified profile is given in Table M2.

The line positions, line widths, sizes and shapes of the emission regions are consistent between the different transitions leaving no doubt about their identification as $^{27}$AlF and $^{26}$AlF. In particular, the consistency in the spatial distribution of the emission assigned to $^{26}$AlF with that of $^{27}$AlF indicates that it must be an isotopologue of AlF – no other species observed in CK Vul has a spatial distribution identical to that of AlF, what is illustrated in Fig. 1. The uncertainty in the calculated positions of the $^{26}$AlF lines of ~1 MHz is insignificant compared to the observed linewidths of approximately 50-120 MHz and the predicted line positions are in excellent agreement with the centers of the emission lines assigned to $^{26}$AlF (in the rest frame of the object). The match between the calculated and observed line positions of $^{26}$AlF are as good as for $^{27}$AlF (Fig. M1), whose rotational spectra were measured in a laboratory. Also, within the observation errors, the line intensities of $^{26}$AlF are consistent with excitation under thermal equilibrium conditions and the excitation temperature of $^{26}$AlF is consistent with that of $^{27}$AlF (see below). In the 7.75 GHz wide spectrum acquired with ALMA in Band 6, we observe only four features of similar intensity and width as these of $^{26}$AlF 7-6. Within the accuracy to which we can determine the observed line positions in CK Vul, the ALMA Band 6 spectrum indicates a probability of a chance coincidence of 1:840. A probability that all four lines match the calculated frequencies is smaller than $10^{-11}$. Therefore, a false identification is highly unlikely.

The spectra acquired to secure the AlF observations serendipitously covered transitions of other species. Additionally, we observed CK Vul with NOEMA in two frequency setups centered at about 89.3 and 146.0 GHz to map emission of species other than AlF. Interferometric imaging of molecular emission was also obtained with the Submillimeter Array (SMA) and reported earlier (*5*). Table M3 contains details of these complementary observations and provides a list of the main lines that were observed. All the interferometric observations allowed us to measure continuum emission. All these extra materials allowed us to trace the complex kinematical, chemical, and excitation structure of the molecular remnant and thus provided an important context for the interpretation of the AlF emission.

All the interferometric data were calibrated using standard procedures. ALMA and NOEMA data were additionally self-calibrated on the strong continuum source. Continuum emission was subtracted from the visibilities as a zeroth- or first order polynomial fitted to the full band and avoiding strong lines. Interferometric maps presented here were obtained in CASA (*35*) and using CLEAN. The weighting of visibilities, natural or robust, was adjusted to the aims of the analysis. We often used images with a restoring beam of a circular shape and of a diameter equal to the geometric mean of the "dirty" beam size.

**Excitation analysis and determination of the isotopic ratio.** The excitation temperature and column densities of the AlF isotopologues were derived in a population-diagram analysis (*37*). The population diagram is shown in Fig. M2. We used a *python*'s *emcee* implementation (*36*) of the Monte Carlo Markov Chains method (*38*) to obtain linear fits to the data. In the associated error analysis, we considered statistical uncertainties from the thermal noise in the flux measurements and a 20% systematic errors in the flux calibrations. We assumed that both isotopologues are located in the same volume and have the same single excitation temperature. That the temperatures are, within uncertainties, equal for both species is evident from the same slopes of lines that can be fitted to both sets of points independently. The source size of 1.80×0.84 arcsec was used to calculate the beam filling factors. This size is a weighted mean of all beam-deconvolved sizes measured in our $^{26}$AlF and $^{27}$AlF maps. Free parameters of the population-diagram fit were the excitation temperature, $T_{\rm ex}$, the column density of $^{27}$AlF, $N_{27}$, and the ratio of the column density of $^{27}$AlF to that of $^{26}$AlF, $N_{27}/N_{26}$. We used uniform ("uninformative") priors for the three parameters allowing their values to be in arbitrary but reasonably broad ranges. A few thousand "walkers" were used in *emcee* to derive the posterior distributions. After the first determination of the column densities, we calculated the line opacities and corrected the measured fluxes for the corresponding saturation (*38*). The calculation of the free parameters was then repeated. The maximum optical thickness in this second iteration was $\tau_{\rm max}$=0.3. The saturation correction is only 1.4% higher than in the previous iteration; that figure is much smaller than uncertainties in the flux measurement and therefore no further corrections were applied. The procedure yielded $T_{\rm ex}=12.9^{+2.4}_{-1.8}$ K, $N_{27}=3.0^{+0.6}_{-0.5}\times10^{15}$ cm$^{-2}$, and $N_{27}/N_{26}=7.1^{+3.2}_{-2.2}$, where the median values are associated with uncertainties corresponding to 97.3% confidence levels. These uncertainties are underestimated and do not take into account, for instance, errors in the source size.

The population-diagram analysis relies on the assumption of thermodynamic equilibrium (TE) in the gas. The assumption is not granted in CK Vul considering that some species appear to be sub-thermally excited (*12*). However, the AlF population diagram itself does not indicate any strong deviations from what is expected in TE. Also, the excitation temperature of AlF derived here is consistent with the kinetic temperature constrained from the single-dish survey (*12*). Using collision rates of AlF with $p$-$H_2$ and with He at 10 K (*39, 40*), we calculated critical densities for all observed transitions. They range from $10^4$ cm$^{-3}$ for $J$=1-0 to $10^7$ cm$^{-3}$ for $J$=7-6 and are therefore comparable to critical densities of analogous transitions of low-density molecular tracers such as CO. The AlF gas is likely thermalized and the level population are likely close to TE.

**Nucleosynthesis of $^{26}$Al.** In order to investigate the synthesis of $^{26}$Al in low-mass stars evolving off the main sequence to RGB, we analyzed state of the art solar-metallicity evolutionary sequences calculated with the Monash stellar-evolution code (*41*). The surface abundances on the asymptotic giant branch (AGB) were investigated with the code in ref. *42*. The models are evolved from the zero-age main sequence to near the end of the thermally-pulsing AGB phase. For the purposes of this study, we sampled the interior composition of the star at the tip of the RGB, before core helium burning is ignited. The grid includes models between 1 and 8 $M_\odot$ and we considered models for 1-3 $M_\odot$ for this study. The nuclear network and initial abundances used for the nucleosynthesis calculations are described in ref. *30* and the input physics used in the stellar evolutionary calculations are described in ref. *42*. Evolution of a grid of low-mass stars (0.9-3.0 $M_{\odot}$) from the pre-main-sequence up to the He-flash (end of the red-giant phase) was also performed using the TYCHO (version 6.0) stellar evolution code developed by David Arnett and his collaborators (*44*).

We find that for all the considered models, the mass of $^{26}$Al increases with the time spent on the RGB. The abundance profiles of the aluminum isotopes at the tip of the RGB (Fig. 2) indicate a high content of $^{26}$Al only in the outermost parts of the well-developed He core. There is not much difference between the amount of $^{26}$Al synthesized in 1 and 2 $M_{\odot}$ models but the total $^{26}$Al mass becomes smaller in models of stars of higher masses. For instance, the total mass of $^{26}$Al in the 1 $M_{\odot}$ model is $9 \cdot 10^{27}$g and only $4 \cdot 10^{26}$ g in the 3 $M_{\odot}$ model. The lower production of $^{26}$Al in the more massive stars is related to their relatively short time spent on the RGB; also, stars with initial masses of $\geq$2.5 $M_{\odot}$ burn H on the main sequence in a convective core which eventually results in lower temperature of H shell burning on the RGB and consequently a lower production of $^{26}$Al. We therefore consider 2.5 $M_{\odot}$ as an upper limit on the mass of the CK Vul's progenitor. Stars with masses $\lesssim$0.8 $M_{\odot}$ evolve longer than the age of the Universe and could not have produced sufficient $^{26}$Al to explain our observations of CK Vul, setting an approximate lower mass limit on the progenitor.

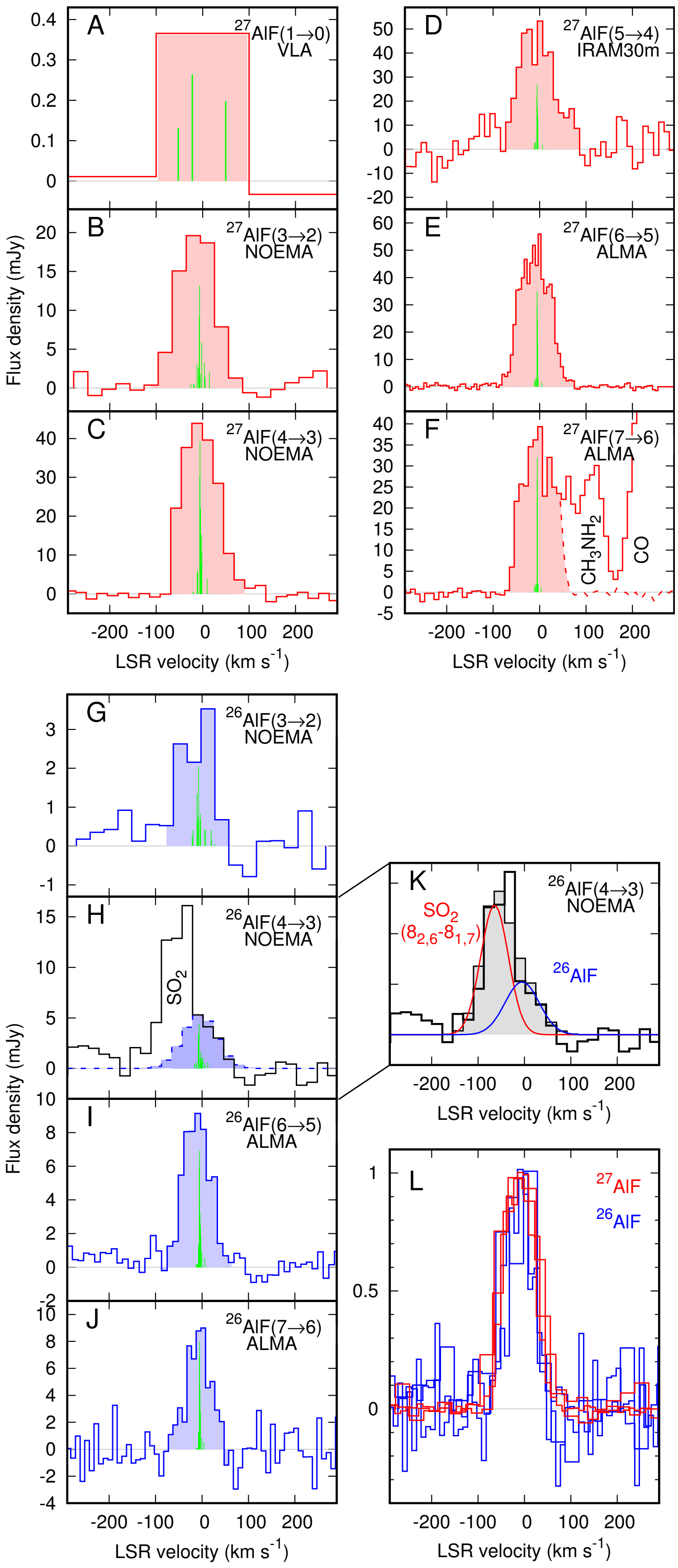

**Fig. M1. Spectra of rotational lines of AlF in CK Vul.** Green vertical lines illustrate the hyperfine structure of the transitions (in arbitrary intensity units). Areas shaded in red and blue show the $^{27}$AlF and $^{26}$AlF emission, respectively, and represent the main emission region of AlF. Some spectra were smoothed in resolution, most heavily for $^{27}$AlF $J$=1-0. The transition and telescope used to collect the data are indicated in each panel. For lines observed with a single-dish telescope and with an interferometer, only the interferometric spectrum is shown. **A-F**, Spectra of $^{27}$AlF. **F**, The $^{27}$AlF 7-6 transition is contaminated by emission of methylamine and a mirrored profile is shown with a dashed line to illustrate the contribution of $^{27}$AlF. **G-J**, Spectra of $^{26}$AlF. **K**, The feature shown with a black empty histogram was decomposed into two Gaussians corresponding to $SO_2$ (red) and $^{26}$AlF (blue). Shaded grey histogram shows the best-fitting combined profile. **L**, Normalized profiles of unblended lines observed with interferometers are overlaid to illustrate their close alignment. Red lines correspond to transitions of $^{27}$AlF, blue ones are $^{26}$AlF.

**Fig. M2. Comparison of AlF emission regions to these of other molecules observed in CK Vul.** Contours are drawn at arbitrarily chosen levels that illustrate the emission morphology. All maps were restored with circular beams for better comparison. **A-B**, Representative maps of emission in $^{26}$AlF and $^{27}$AlF. **C-H**, Sample maps of emission in species displaying a wide range of morphologies. **C**, Emission in $SO_2$ is most similar to that of AlF but has a different position angle of its major axis. **D-E**, The emission of the only two molecular ions observed in this source appear generally more extended than AlF. **F-G**, In contrast to AlF, most of the observed species have a strong component within the bipolar lobes.

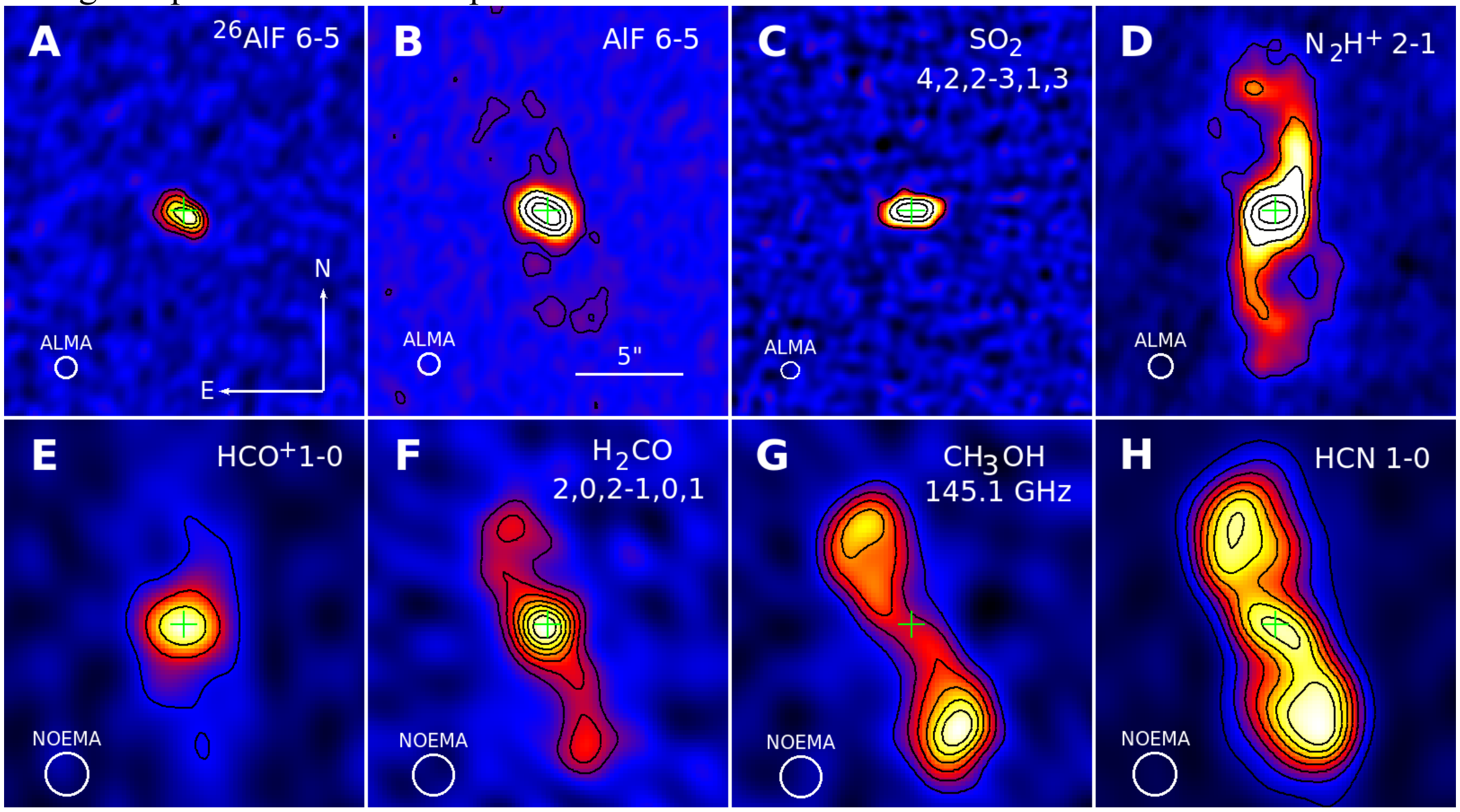

**Fig. M3. The population diagram for AlF in CK Vul.** Shaded cyan and green areas are 2000 trial Monte Carlo fits to the data. The errorbars represent only statistical errors.

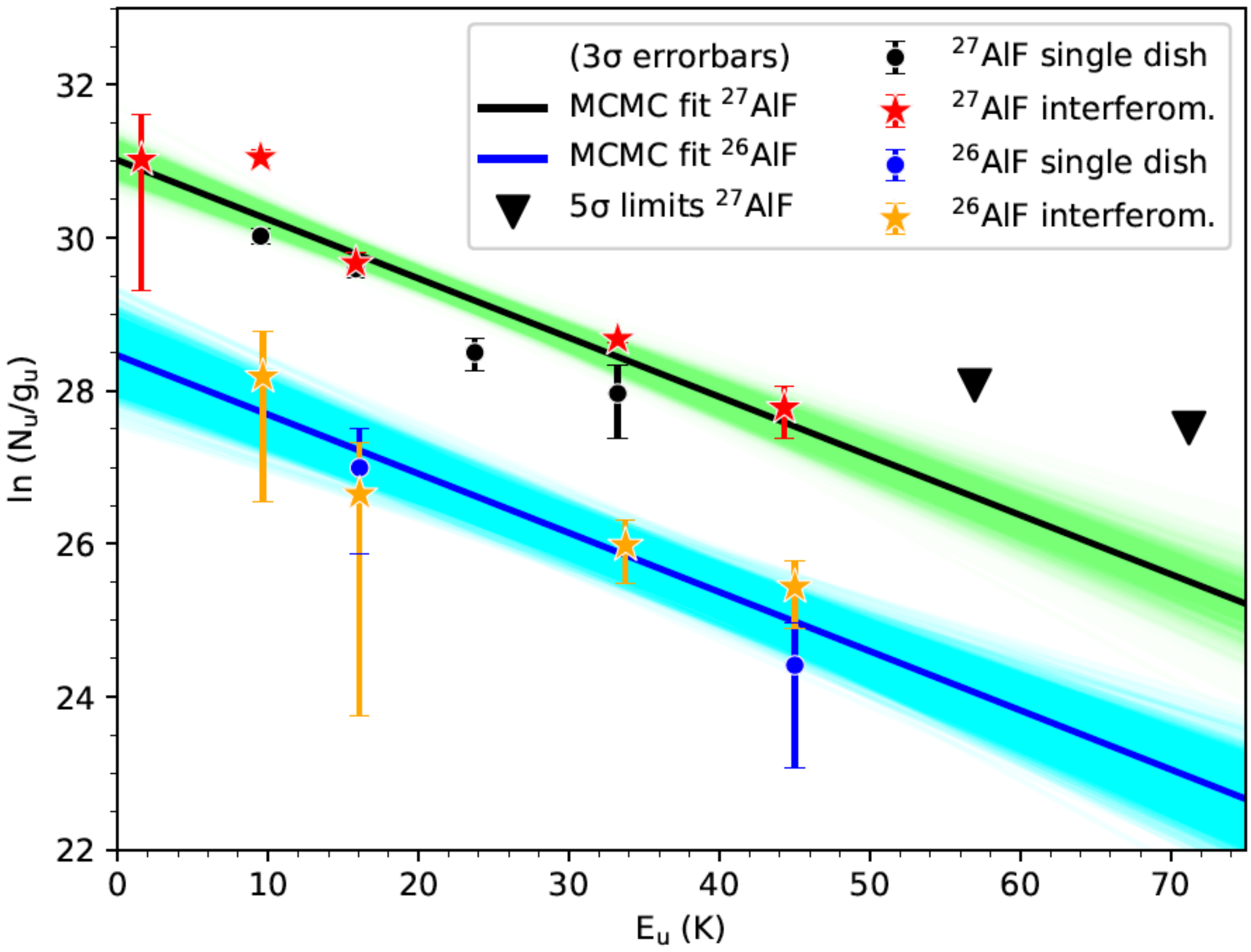

**Table M1.** Spectroscopic constants in MHz of $^{26,27}$AlF used in our calculations. The 1σ errors are given in brackets.

| | $^{27}Al^{19}F$ | $^{26}Al^{19}F$ |
|---|---|---|
| $B \times 10^{-4}$ | 1.64883599(17) | 1.67485239(18) |
| $D \times 10^{2}$ | 3.1398(59) | 3.2399(61) |
| $H \times 10^{9}$ | -9.14(28) | -9.58(29) |
| $eQq_0$(Al) | -37.53(9) | -67.8(83) |
| $c_I$(Al) $\times 10^3$ | 8(6) | 3(2) |

**Table M2.** Spectroscopic data and flux measurements of AlF features.

| **Line ($J_u$-$J_l$)** | **Frequency (GHz)** | **$E_u$ (K)** | **$S\mu_{ul}^2$ (D)** | **Beam (")** | **Flux (Jy km/s)** | **Flux err. (Jy km/s)** | **Instrument** |
|---|---|---|---|---|---|---|---|
| **$^{27}$AlF single dish** | | | | | | | |
| 3-2 | 98.92674 | 9.50 | 41.78 | 24.9 | 2.90 | 0.10 | IRAM 30m |
| 4-3 | 131.89879 | 15.83 | 56.46 | 18.7 | 5.90 | 0.21 | IRAM 30m |
| 5-4 | 164.86782 | 23.74 | 70.38 | 14.9 | 4.85 | 0.34 | IRAM 30m |
| 6-5 | 197.83306 | 33.23 | 84.29 | 31.5 | 6.08 | 0.90 | APEX |
| 8-7 | 263.74919 | 56.97 | 113.16 | 23.7 | <23.84† | 4.77 | APEX |
| 9-8 | 296.69855 | 71.21 | 126.81 | 21.0 | <21.75† | 4.35 | APEX |
| **$^{27}$AlF interferometric** | | | | | | | |
| 1-0 | 32.97659 | 1.58 | 14.23 | 1.3 | 0.058@ | 0.016 | JVLA |
| 3-2 | 98.92674 | 9.50 | 41.78 | 0.8 | 2.33 | 0.08 | NOEMA |
| 4-3 | 131.89879 | 15.83 | 56.46 | 1.7 | 4.22 | 0.20 | NOEMA |
| 6-5 | 197.83306 | 33.23 | 84.29 | 1.0 | 4.94 | 0.09 | ALMA |
| 7-6 | 230.79377 | 44.31 | 97.94 | 0.8 | 2.87‡ | 0.32 | ALMA |
| **$^{26}$AlF single dish** | | | | | | | |
| 3-2 | 100.48765 | 9.65 | 77.4 | 24.5 | <0.39# | 0.13 | IRAM 30m |
| 4-3 | 133.97990 | 16.08 | 103.38 | 18.4 | <0.95§ | 0.22 | IRAM 30m |
| 6-5 | 200.95430 | 33.76 | 153.94 | 31.0 | <4.51† | 0.90 | APEX |
| 7-6 | 234.43488 | 45.01 | 178.98 | 10.5 | 0.67 | 0.17 | IRAM 30m |
| **$^{26}$AlF interferometric** | | | | | | | |
| 3-2 | 100.48765 | 9.65 | 77.4 | 0.8 | 0.29 | 0.08 | NOEMA |
| 4-3 | 133.97990 | 16.08 | 103.38 | 2.0 | 0.49§¶ | 0.15 | NOEMA |
| 6-5 | 200.95430 | 33.76 | 153.94 | 1.0 | 0.70 | 0.09 | ALMA |
| 7-6 | 234.43488 | 45.01 | 178.98 | 0.8 | 0.56 | 0.08 | ALMA |

**Notes:** Geometric mean of the beam size is given. Flux errors are 1$\sigma$ noise uncertainties. † 5$\sigma$ upper limits. @ Tentatively detected. ‡ Feature may be slightly contaminated by CO 2-1 emission and partially blends with a transition of $CH_3NH_2$. Flux was derived by mirroring the unaffected part of the profile. # Contaminated by an instrumental artifact. § Contaminated by emission of $SO_2$. ¶ Flux derived in a deblending procedure.

**Table M3.** Interferometric observations of AlF in CK Vul.

| Array | Band | Observation dates | Ang. resolution ("), beam PA (°) | No. of antennas or pads | Frequency range (GHz) | Native spectral binning (kHz) | Main lines observed |
|---|---|---|---|---|---|---|---|
| JVLA | *Ka* | 18, 22 Jan 2016 | 1.5×0.7, 85.9 | 27 | 29.9-37.9 | 1000 | $HC^{13}CCN$+$HCC^{13}CN$ (4-3), $HC_3N$ (4-3), $CH_3CN$ (2,0-1,0), **AlF (1-0)** |
| NOEMA | Band 1 | 24 Sep 2016 | 2.7×1.5, 15.5 | 8 | 87.4-91.1 | 10000 | HNCO(4,0,4-3,0,3), HCN(1-0), $H^{15}NC$(1-0), $HCO^+$(1-0), HNC (1-0), SiS(5-4), $HC_3N$(10-9) |
| NOEMA | Band 1 | 26, 28 Jan 2016; 16, 18 Jan 2017 | 1.0×1.6, 34.9 | 7, 8 | 98.2-101.8 | 10000 | SO(2,3-1,2), $HC^{13}CCN$+ +$HCC^{13}CN$ (11-10), $HC_3N$ (11-10), $SO_2$(2,2,0-3,1,3), **AlF(3-2)**, **$^{26}$AlF(3-2)** |
| NOEMA | Band 2 | 18 Nov 2015 | 2.0×1.5, 16.0 | 6 | 130.9-134.5 | 10000 | $HC^{13}CCN$(15-14), $CH_3NH_2$ 133.98 GHz, **AlF(4-3)**, **$^{26}$AlF(4-3)** |
| NOEMA | Band 2 | 27 Jul 2016 | 2.1×1.7, 41.3 | 8 | 144.2-147.8 | 10000 | $C^{34}S$ (4-3), $CH_3OH$ (3,$K_a$,$K_c$-2,$K_a$,$K_c$), $H_2CO$ (2,0,2-1,0,1), $H_2{}^{13}CO$ (2,1,1-1,1,0), CS (3-2), $CH_3CN$ (8-7) |
| ALMA | Band 5 | 4, 5 Apr 2018 | 1.0×0.9, 156.5 | 41, 43 | 185.7-189.4; 197.7-201.4 | 3906.3 | $N_2H^+$(2-1), SO(6,6-6,7), PN(4-3), $SO_2$(9,2,8-9,1,9), $^{13}CH_3OH$ 188.8 GHz, **AlF(6-5)**, SiS(11-10), $CH_2NH$(3,1,2,2-2,1,1,2), $HC_3N$(22-21), **$^{26}$AlF(6-5)** |
| ALMA | Band 6 | 2, 3 Jun 2016 | 0.8×0.8, 96.5 | 41 | 230.6-232.4; 233.5-235.4; 246.1-248.0; 248.1-250.0 | 3906.3 | CO (2-1), $CH_3NH_2$ (3,2,4-3,1,5), $^{13}CS$ (5-4), PN (5-4), $SO_2$ (4,2,2-3,1,3), $^{34}SO$ (6,5-5,4), **AlF(7-6)**, **$^{26}$AlF(7-6)** |
| SMA | 345 | 3 Jul 2014 | 2.0×1.3, 270 | 8 | 330.2-332.2; 335.2-337.2; 345.2-347.2; 350.2-352.2 | 640.4 | $^{13}CO$(3-2), $H^{13}CN$(4-3), CO(3-2), $H^{13}CO^+$(4-3) |
| SMA | 230 | 30 Jul 2014 | 8.4×4.7, 71.5 | 7 | 216.9-218.8; 218.9-220.8; 228.9-230.8; 230.9-232.8 | 640.4 | SiO(3-2), $^{13}CN$(2-1), $C^{18}O$(2-1), $^{13}CO$(2-1), $^{13}CS$(5-4) |

**Notes:** Angular resolution is given for robust weighting of visibilities.